\documentclass[a4paper]{llncs}

\usepackage{graphicx}
\usepackage{vdmlisting}
\usepackage{listing}
\usepackage{hyperref}
\usepackage{url}
\lstdefinestyle{overtureLanguageStyle}{basicstyle=\footnotesize\ttfamily,
			numbers=left,
                        frame=trBL, 
                        tabsize=2, 
                        linewidth=\textwidth,
                        showstringspaces=false, 
                        captionpos=b,
                        frameround=fttt, 
                        aboveskip=2mm,
                        belowskip=2mm,
                        framexleftmargin=0mm, 
                        framexrightmargin=0mm,
                        escapeinside={(*@}{@*)},
                        language=VDM_SL}
\lstdefinelanguage{python}{ 
  backgroundcolor={\color[gray]{1}},
  basicstyle=\small\ttfamily,
  sensitive=true, 
  morestring=[s]{"}{"}, 
} 
\lstdefinelanguage{vdmsl}{ 
  backgroundcolor={\color[gray]{1}},
  basicstyle=\small\ttfamily,
  sensitive=true, 
  morestring=[s]{"}{"}, 
} 

\title{Trace-Based Execution-Level Observability of VDM-SL Specifications}

\titlerunning{24rd Overture Workshop, 2026}

\author{
Tomohiro Oda\inst{1} \and 
Han-Myung Chang\inst{2}
}

\authorrunning{Oda, T.; Chang, H.M.}

\institute{Software Research Associates, Inc.~(\email{tomohiro@sra.co.jp})
\and Nanzan University~(\email{chang@nanzan-u.ac.jp}) 
}
\begin{document}

\maketitle

\begin{abstract}
VDM has been pursuing rigorous verification through mathematical theorem proving and software testing via simulated execution. 
Animation through an interpreter enables validation of the specification to ensure it meets the required functionality.
Step-by-step execution in a debugger also allows the user to follow the internal behavior of operations.
In this paper, we propose the recording and utilization of execution traces of assignments, operation calls, and return statements to make the internal behavior of operations persistent and analyzable as state-based models.
The data model of events in execution traces, its implementation in ViennaTalk, and its application to visualization will be introduced.
\end{abstract}

%%%%%%%%%%%%%%%%%%%%%%%%%%%%%%%%%%
\section{Introduction}

Model-based formal specification languages describe systems through abstract models expressed in a notation with a mathematical foundation.
VDM-SL follows this tradition: a specification defines a state space and a set of operations that refer to and may transition between states~\cite{VDM10LRM}.

The executable subset of VDM-SL allows developers to simulate the behavior of the specified system through interpreted execution.
This executability enables developers to explore system behavior early in the specification process through animation, testing~\cite{CombinatorialTesting,VDMUnit}, and scenario-based experimentation~\cite{WiderRangeOfStakeholders}.
In exploratory specification, animation serves as an effective means for specificatin engineers to deepen their understanding of the system under development through iterative execution and inspection.
It also functions as a boundary object that facilitates communication and shared understanding between formal engineers and stakeholders who may not be familiar with formal methods.

Assertions play important roles in both formal verification methods and animation-based techniques.
For example, state invariants in VDM can be checked at every single assignment, whereas in typical test frameworks for programming languages, assertions are hand-written in the test code and can only check the state before and after the execution of the target code.
Preconditions and postconditions allow for checking the arguments, return values and states of operations invoked by nested calls.

Assertions thus give opportunities to check expected properties at discrete step of the execution.
On the other hand, properties spanning multiple execution steps are not in the scope of assertions in VDM.
Invariants and preconditions may only refer to the state of the moment, and postconditions may also refer to the states before and after the execution of the operation.
Other formalisms, such as CSP, provide theory and tools to handle those properties that span multiple events in possible execution paths.

In this paper, we complement assertions of VDM with microscopic observations of individual executions.
We propose a trace-based approach to execution-level observability of VDM-SL specifications, in which execution traces, consisting of operation calls, returns, and state updates, are captured and systematically analyzed.
Rather than characterizing all possible behaviors, we concentrate on making concrete executions persistent and inspectable artifacts that support behavioral inspection and visualization.

The remainder of this paper is organized as follows. Section 2 defines the data structure of execution traces for VDM-SL. Section 3 describes the implementation in ViennaTalk and its visualization features. Section 4 discusses related work, and Section 5 concludes the paper.

\section{Execution Trace for VDM-SL}
\label{sec:execution-trace}

To realize the trace-based observability discussed in the previous section, we first define what information should be captured from execution of VDM specifications.
VDM is a state-based modeling technique.
Each module in VDM-SL defines a state space through a state section, alongside four other kinds of sections: types, operations, values and functions.
A state space is composed of one or more state variables with static types and initial values.
The operations section defines a set of operations that may observe and also may modify the state of a module.
An operation has its type signature, and its body may refer to constants defined in values sections and also may apply functions.

\begin{figure}
\begin{vdmsl}
types
    ExecutionTrace = RootEvent;
    RootEvent :: 
        initial : set of AssignEvent events : seq of Event;
    Event = AssignEvent | OpCallEvent | OpReturnEvent;
    AssignEvent :: modName : Name assigns : map Name to ?;
    OpCallEvent ::
        modName : Name
        opName : Name
        arguments : seq of ?
        events : seq1 of Event;
    OpReturnEvent :: val : ?;
    Name = seq1 of char;
\end{vdmsl}
\caption{Definition of Execution Trace}
\label{fig:definition-of-execution-trace}
\end{figure}

From the perspective of stateful modeling, each module can be viewed as a state machine, and the language constructs associated with states are of primary focus for observation.
A state variable is accessible only within the module, and therefore states can be observed and/or transitioned exclusively through variable references and assignments in operations.
The states of the modules in a specification are coordinated through inter-module operation calls.
To capture the execution-level dynamics of the specified system, we focus on tracing assignment statements to state variables and operation calls.

Fig. \ref{fig:definition-of-execution-trace} describes the execution trace.
An execution trace is defined as a tree structure rooted at a root event (Line 2).
The root event consists of the initial state of the traced execution and a sequence of events captured directly in the execution (Line 3--4).
An event is either an assignment event, an operation call event, or an operation return event (Line 5).

An assignment event records the module name and a mapping of variable names to their assigned values (Line 6).
The assignment event can hold multiple assignment information to accomodate the atomic assignment statements in VDM-SL.
Since  it has no sub-events, the assignment event is a leaf node in the tree structure.

An operation call event has the module name and the operation name to record invocation of the operation, and also holds events captured in the execution of the callee operation (Line 7--11).
An operation return event has the module name, the operation name and the returned value (Line 12).
Please note that the question marks that appear in place of types denote the {\it any} type.
The {\it any} type informally represents an unconstrained type that can hold values of any kind, effectively acting as a top type in the type system.
Although the {\it any} type is not a legitimate construct in VDM-SL, it is accepted by some tools including VDMJ and ViennaTalk.

\subsection{Example: Automatic Door}

\begin{figure}
\begin{vdmsl}
module AD
exports all
definitions
types
    Presence = <PRESENT> | <VACANT>;
    Position = <CLOSED> | <PARTIAL> | <OPEN>;
    Time = nat;
    Command = <OPEN> | <CLOSE> | <HALT>;
values
    OPEN_DURATION : Time = 2;
state AD of
    personPresence : Presence
    doorPosition : Position
    timeToClose : Time
    command : Command
inv mk_AD(p, d, t, c) == t > 0 => d = <OPEN> and t <= 2
init s == s = mk_AD(<VACANT>, <PARTIAL>, 0, <HALT>)
end
operations
    tick : () ==> ()
    tick() == (checkPersonPresence(); moveDoor());
    checkPersonPresence : () ==> ()
    checkPersonPresence() ==
        if personPresence = <PRESENT> then command := <OPEN>;
    moveDoor : () ==> ()
    moveDoor() ==
        cases command:
            <OPEN> -> openDoor(),
            <CLOSE> -> closeDoor(),
            <HALT> -> 
                if doorPosition = <OPEN> then tickCloseTimer()
            end;
    openDoor : () ==> ()
    openDoor() ==
        cases doorPosition:
            <CLOSED> -> doorPosition := <PARTIAL>,
            <PARTIAL> -> doorPosition := <OPEN>,
            <OPEN> -> (command := <HALT>; 
                       timeToClose := OPEN_DURATION)
            end;
    closeDoor : () ==> ()
    closeDoor() ==
        cases doorPosition:
            <CLOSED> -> command := <HALT>,
            <PARTIAL> -> doorPosition := <CLOSED>,
            <OPEN> -> doorPosition := <PARTIAL>
            end;
    tickCloseTimer : () ==> ()
    tickCloseTimer() ==
        if timeToClose > 0 
        then timeToClose := timeToClose - 1 
        else command := <CLOSE>;
end AD
\end{vdmsl}
\caption{Specification of Automatic Door}
\label{fig:automatic-door-source}
\end{figure}

We introduce an example specification of a simple automatic door in Fig.~\ref{fig:automatic-door-source}.
The door is equipped with a sensor to detect the presence of a person.
If a person is detected, the door should open, remain open for a certain duration if the door becomes vacant, and then begin closing.
This specification assumes that the door controller has a clock signal generator that periodically invokes the \texttt{tick} operation.

The automatic door has four state variables, as declared in Fig.~\ref{fig:automatic-door-source}.
The \texttt{person\-Presence} variable indicates whether a person is present (\texttt{<PRESENT>}) or the area is vacant (\texttt{<VACANT>}) (Lines 5 and 12).
The \texttt{doorPosition} variable represents the current state of the door: closed, partially open, or fully open (Lines 6 and 13).
The \texttt{timeToClose} variable serves as a count-down timer to keep the door open (Lines 7 and 14), while \texttt{command} stores the current instruction for the motor: open, close, or halt (Line 8 and 15).
Based on the state invariant asserted in Line 16, the total number of legal states is 20.

The \texttt{tick} operation, defined in Lines 20 and 21 of Fig.~\ref{fig:automatic-door-source}, first calls the \texttt{check\-Person\-Presence} operation. 
This operation commands the motor to open the door if a person is detected. 
Subsequently, the \texttt{moveDoor} operation is called to control the actual movement of the door.
The operations \texttt{openDoor}, \texttt{closeDoor}, and \texttt{tick\-CloseTimer} are invoked by \texttt{moveDoor} according to the value of the \texttt{command} variable.

\begin{figure}
\begin{vdmsl}
mk_RootEvent({mk_AssignEvent("AD", 
   {"personPresence" |-> <VACANT>, "doorPosition" |-> <PARTIAL>,
    "timeToClose" |-> 0, "command" |-> <HALT>})}, 
    [mk_AssignEvent("AD", {"personPresence" |-> <PRESENT>}),
    mk_OpCallEvent("AD", "tick", [],
        [mk_OpCallEvent("AD", "checkPersonPresence", [],
            [mk_AssignEvent("AD", {"command" |-> <OPEN>}),
            mk_OpReturnEvent(nil)]),
        mk_OpCallEvent("AD", "moveDoor", [],
            [mk_OpCallEvent("AD", "openDoor", [],
                [mk_OpReturnEvent(nil)]),
            mk_OpReturnEvent(nil)])])])
\end{vdmsl}
\caption{An example execution trace of Automatic Door}
\label{fig:automatic-door-execution-trace}
\end{figure}

Fig.~\ref{fig:automatic-door-execution-trace} shows the execution trace resulting from the execution of the following statement:
\begin{vdmsl}
(personPresence := <PRESENT>; tick(); personPresence := <VACANT>)
\end{vdmsl}

The execution trace in Fig.~\ref{fig:automatic-door-execution-trace} reveals all states reached during execution; for each state transition, we can identify the operations directly or indirectly involved.
A call tree can also be constructed from this execution trace.
This event tree can also be transformed into a linear sequence by traversing the nodes in a depth-first search (DFS) manner.
We chose a tree data structure to accommodate features such as exception handling, which can complicate the reconstruction of a hierarchical structure from a flat sequence of events.

% maybe it's a good idea to provide an example animation and its execution trace.

\section{Implementation in ViennaTalk}

ViennaTalk is an IDE for VDM-SL designed to support the early stages of the formal specification phase, where frequent modification to the tentative specification and evaluation based on the simulated execution are conducted\cite{ViennaTalk}.
In such exploratory phase, understanding the implications of the model and collecting feedback from stakeholders, such as domain experts, is important.
We implemented the execution trace described in Section \ref{sec:execution-trace} to enable finer inspection of the simulated executions.
In this section, we explain how the execution trace is implemented in ViennaTalk and explain functionalities leveraging the execution trace.

\subsection{ViennaTalk and Pharo}

ViennaTalk is implemented on top of Pharo\cite{Pharo}, which provides the reflection and meta-programming capabilities useful for hosting guest languages.
Pharo is a successor dialect to Smalltalk, featuring a modern and modular implementation.
One highlight of Pharo is immersive development environment: the bytecode of methods is executed in the same process as the IDE itself.
The running code can refer to the IDE and the IDE can inspect everything in the running code including its stack frame in the virtual machine, allowing the programmer to dive into the live objects in the execution.
This immersiveness fits well with ViennaTalk’s exploratory style.

\begin{figure}
\begin{center}
\includegraphics[width=0.8\textwidth]{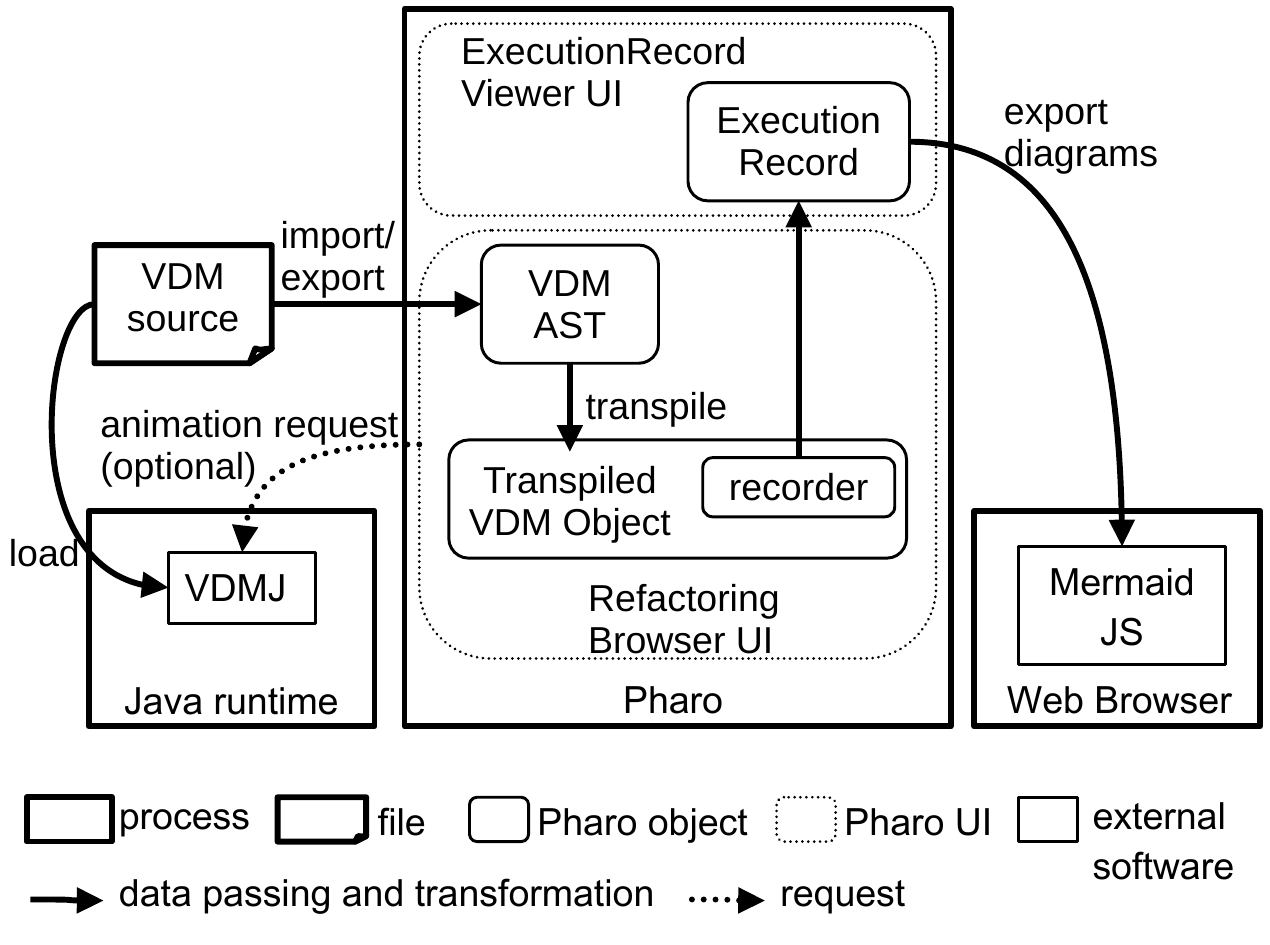}
\end{center}
\caption{An overview of ViennaTalk's execution trace and related tools}
\label{fig:architecture}
\end{figure}

Fig \ref{fig:architecture} shows the overview of ViennaTalk's components for execution traces and associated tools.
To take advantage of the Pharo's immersive programming environment, ViennaTalk transpiles a specification in VDM-SL into Pharo classes and methods (noted as Transpiled VDM Object in Fig \ref{fig:architecture}), and runs them as Pharo code so that the immersive tools in Pharo can be used to develop VDM-SL specifications \cite{CodeGenerator} while interpretation via VDMJ is optionally supported.
Consequently, ViennaTalk's execution trace mechanism is integrated directly into the transpiler, ensuring that the generated Pharo code automatically records its trace during execution.
Based on this infrastructure, Pharo-based tools have been implemented to support the user's development of VDM-SL specifications, including an interactive viewer for hierarchical execution traces, UML-based visualizations, and functionality to export/import traces as CSV files.

Another important feature of Pharo is the {\it slot} mechanism, which allows programs to control what bytecode is emitted on read and write access to a variable.
In Pharo, a slot can be defined by subclassing the {\tt Slot} class, and an instance variable can be declared with a specific slot class.
ViennaTalk uses this mechanism to handle state invariants.
State variables with invariants are declared using the {\tt Vienna\-State\-Variable\-With\-Inv\-Slot} class, which generates bytecode to evaluate the invariant on write access, whereas variables without invariants are declared using the {\tt Vienna\-State\-Variable\-Slot} class.
This mechanism is also used to record assignments during execution, as described in Section~\ref{sec:capturing-execution-trace}.

\subsection{Capturing Execution Trace}
\label{sec:capturing-execution-trace}

\begin{figure}
\begin{center}
\includegraphics[width=1\textwidth]{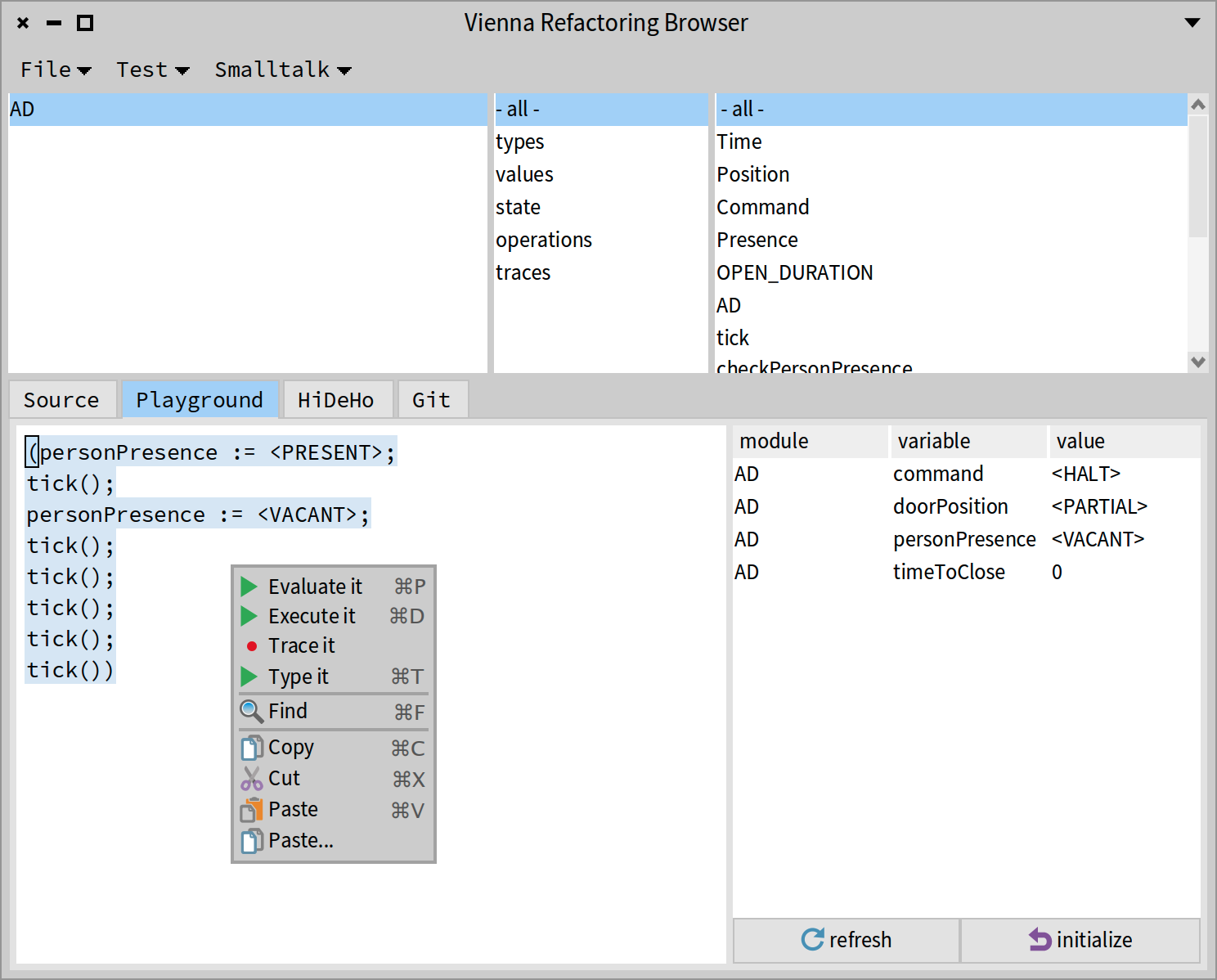}
\end{center}
\caption{A screenshot of Playground on Refactoring Browser}
\label{fig:playground-trace-it}
\end{figure}

ViennaTalk is equipped with the Vienna Refactoring Browser to facilitate the development of VDM-SL specifications\cite{RefactoringBrowser}.
The Refactoring Browser is more than a simple text editor; it allows users to write and execute snippets of VDM expressions or statements on its Playground page.

Fig.~\ref{fig:playground-trace-it} shows a screenshot of the Vienna Refactoring Browser.
The browser window is divided into an upper and a lower pane.
The upper pane contains three lists for navigating the hierarchical structure of the specification: modules, sections, and top-level definitions.
In Fig.~\ref{fig:playground-trace-it}, the {\tt AD} module is selected.
The lower pane contains four tabs in a tabbed interface.
In Fig.~\ref{fig:playground-trace-it}, the {\tt Playground} tab is selected.
The {\tt Playground} tab contains a text editor on the left and a table displaying the current system state on the right.
In the text editor, the user can write arbitrary expressions and statements.
The user can select a portion of the text and then invoke commands for execution, type inspection, and editing from the context menu.

The snippet shown in Fig.~\ref{fig:playground-trace-it} simulates a scenario in which a person appears while the door is partially open.
In this scenario, the \texttt{tick} operation is executed, after which the person leaves.
The \texttt{tick} operation is then executed five times in succession.
Executing this snippet causes the door to begin closing while remaining in a partially open state.
The {\tt Trace it} command generates Pharo classes and methods instrumented with probes to record assignments and operation calls and returns.

The specification object transpiled from the VDM specification maintains a recorder object that stores a sequence of events in a tree structure.
The probes embedded in the transpiled code send requests to this recorder object.

For assignments, the {\tt Vienna\-State\-Variable\-Slot\-For\-Execution\-Tracing} class inserts code that sends a request to the recorder on each write access.
For operation calls, the transpiler inserts code to record the call together with its arguments.
For operation returns, the transpiler sets up unwinding handlers to record the return event along with the return value.
Once executed, the collected trace is presented in a dedicated viewer, which is described in the next section.

\subsection{Execution Trace Viewer}

\begin{figure}
\begin{center}
\includegraphics[width=1\textwidth]{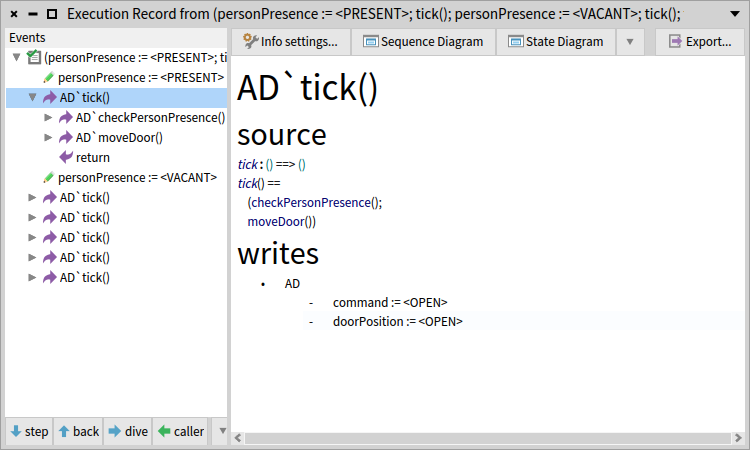}
\end{center}
\caption{A screenshot of execution trace viewer}
\label{fig:trace-viewer}
\end{figure}

The Execution Trace Viewer provides a graphical user interface (GUI) for navigating and inspecting recorded execution traces. Fig.~\ref{fig:trace-viewer} shows a screenshot of the viewer displaying the trace captured from the snippet in Fig.~\ref{fig:playground-trace-it}.

The left pane displays the execution trace in a hierarchical tree format. 
Distinct icons represent different types of events: the root event is marked with a notebook-like icon, while assignment events are indicated by pencil icons. 
Operation call events are denoted by magenta right arrows. 
Each operation call event includes a small triangle that allows the user to toggle the expansion or collapse of its nested sub-events. 
Return events are represented by magenta left arrows.

When an event is selected in the tree, its details are displayed in the right pane.
In Fig.~\ref{fig:trace-viewer}, the operation call for \texttt{tick} is selected.
The right pane shows the call with its actual arguments (in this case, no arguments are passed), the corresponding VDM source definition, and a summary of write accesses performed during the operation.

The viewer also functions as a simple time-travel debugger.
Four navigation buttons are located beneath the event tree.
The \textit{Step} button moves to the next event at the same hierarchical level, similar to a step-over operation in a conventional debugger, while the \textit{Back} button performs the reverse.
The \textit{Dive} button allows the user to step into the callee's context when an operation call is selected.
Conversely, the \textit{Caller} button moves the selection to the caller's context.
These navigation features enable users to efficiently locate the point of failure and trace back to its root cause.

Above the description pane, the user can visualize the specific event selected in the left tree view. 
Typically, the user might view a sequence diagram or a state diagram of the root event to capture a high-level overview of the system's dynamic behavior. 
Additionally, the viewer allows for localized visualization, making it possible to generate diagrams confined to a selected operation call event.

Above the description pane, the user can visualize the event selected in the left tree view.
The user can typically view a sequence diagram and a state diagram of the root event to capture the overview of the dynamic behavior of the system, it's also possible to visualize only within a selected operation call event.

\subsection{Sequence Diagram Visualization via mermaid}

\begin{figure}
\begin{center}
\includegraphics[width=1\textwidth]{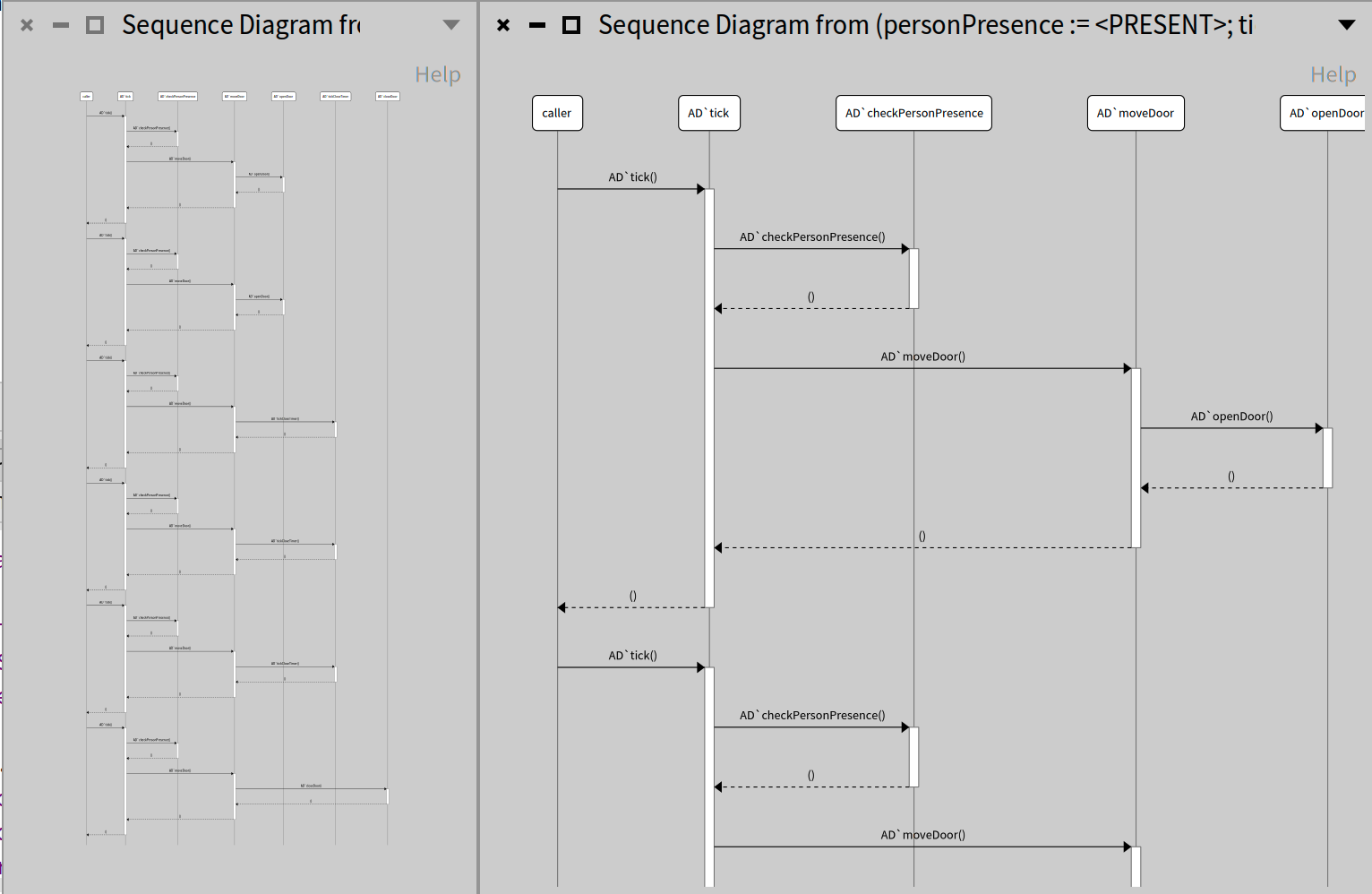}
\end{center}
\caption{A screenshot of a sequence diagram generated from execution trace}
\label{fig:sequence-diagram}
\end{figure}

ViennaTalk provides a functionality to transform an execution trace into a sequence diagram-like visualization.
Fig. \ref{fig:sequence-diagram} shows two screenshots viewing the same diagram; the left pane presents the panned overview of the diagram and the right pane from a magnified view.
Through this visualization, the user can follow the flow of operation calls with actual arguments, nested operation calls, and the return values.
The vertical lines with rounded-rectangle headers represent operations, whereas they are objects and actors in the standard sequence diagrams in UML.
The expression or statement that was subject to the \textit{Trace it} operation is labeled {\tt caller}.

The horizontal arrows in solid lines indicate operation calls, printing the operation name and arguments above it.
The callee operations are activated until the control is returned to the caller operations, rendered as horizontal arrows in broken lines with the return values printed above them.
Activations are denoted as strings.

The visualization is generated using the Mermaid notation\cite{Mermaid}.
Users can save the visualization as a Mermaid script and embed it directly into Markdown documents.
Since Pharo provides a Markdown renderer in its standard library, users can include these Mermaid diagrams within class or package comments as documentation.
Additionally, the generated Mermaid scripts can be embedded into a \texttt{README.md} file in a GitHub repository to enhance the project's external documentation.

\subsection{State Diagram Visualization via mermaid}

\begin{figure}
\begin{center}
\includegraphics[width=1\textwidth]{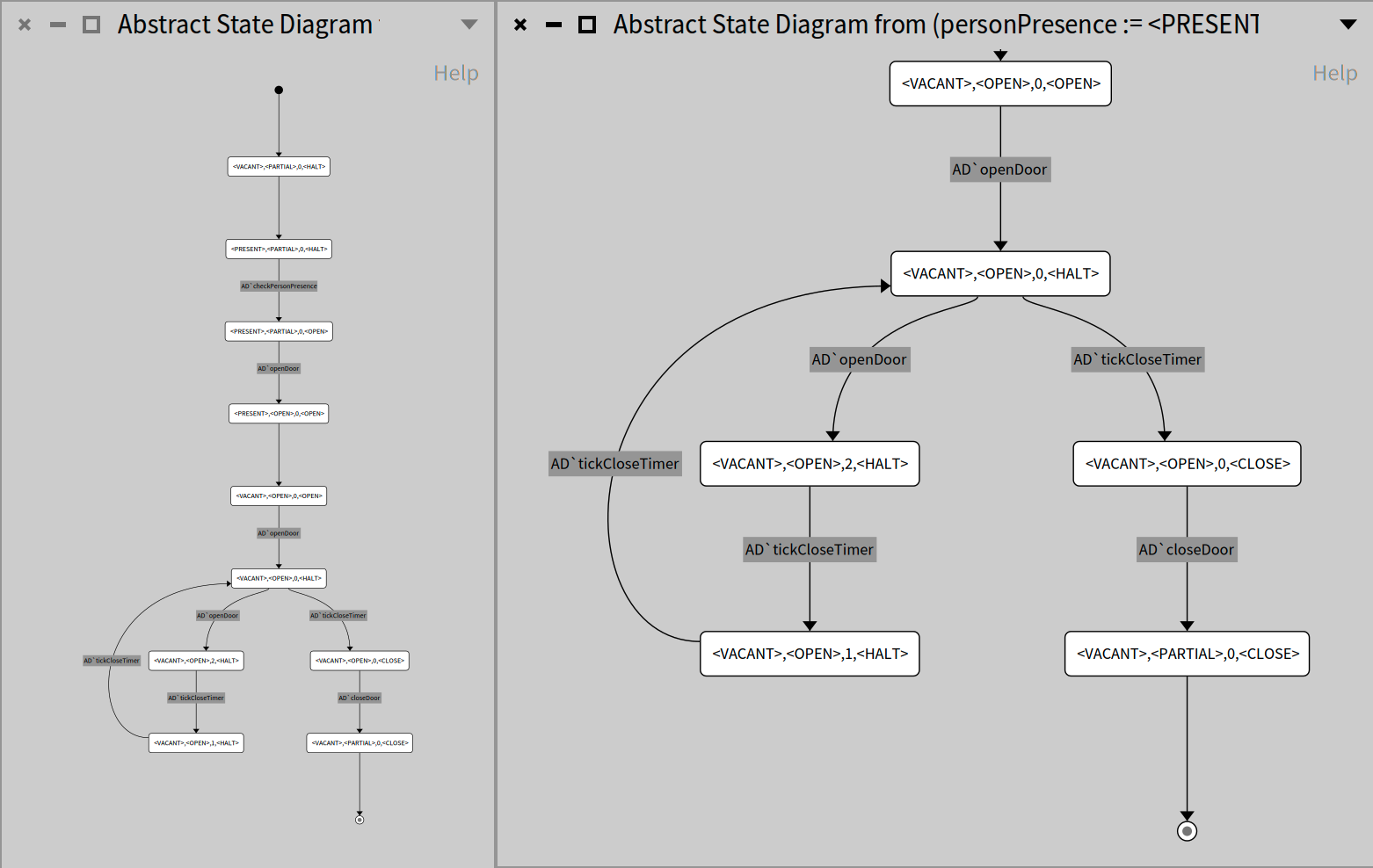}
\end{center}
\caption{A screenshot of a state diagram generated from execution trace}
\label{fig:state-diagram}
\end{figure}

State diagrams can also be generated from execution traces via mermaid script.
While a sequence diagram illustrates the flow of control between operations, a state diagram visualizes the transitions between system states and identifies which operation triggered the transition.

Fig. \ref{fig:state-diagram}  shows two screenshots viewing the same state diagram: the overview on the left and the zoomed view on the right.
This diagram was generated by executing of the snippet shown in Fig. \ref{fig:playground-trace-it}.
In this execution, the sensor first detects a person, triggering the door to open. Once fully opened, the door remains at a halt for three ticks before it begins closing.

In the zoomed view, the state \texttt{<VACANT>,<OPEN>,0,<HALT>} forms a loop, indicating that the system passed through the same state twice.
Although this repeated state does not necessarily imply a defect, it can be confusing or misleading when observing the execution process, as the intended behavior does not involve repeating the same action.

The state diagram in Fig. \ref{fig:state-diagram} helps identifying the cause of this reoccurrent state.
The first transition to this state was performed by the \texttt{AD`openDoor} operation, which then transitioned the system to the \texttt{<VACANT>,<OPEN>,2,<HALT>} state.
These two transitions resulted from executing the Line 38--39 in Fig. \ref{fig:automatic-door-source}.
Since the two assignment statements were executed in series, an intermediate state was produced between them, leading to the confusing recurrence.

\begin{figure}
\begin{center}
\includegraphics[width=0.7\textwidth]{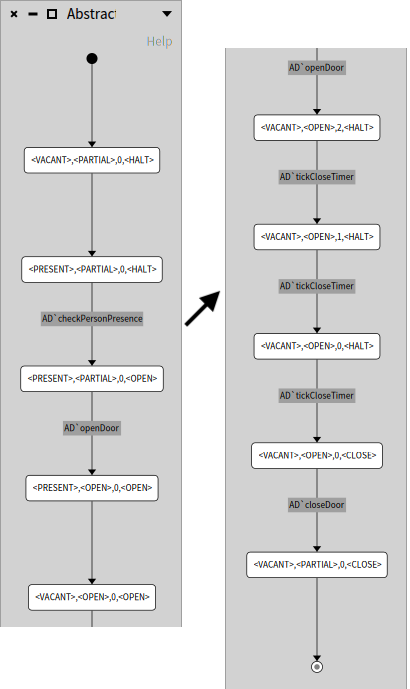}
\end{center}
\caption{A screenshot of a state diagram generated from execution trace}
\label{fig:state-diagram-fixed}
\end{figure}

Encapsulating these two assignments into a multiple assignment statement using the \texttt{atomic} keyword resolves the issue.
A multiple assignment statement treats assignments to multiple variables as a single state transition.
If an invariant is defined on the state, it is not checked between individual assignments; instead, it is evaluated only after all assignments have been completed.
Consequently, in the execution trace, the entire sequence is recorded as a single state-change event.
Fig.~\ref{fig:state-diagram-fixed} shows the state diagram obtained from the same snippet with the modified specification.
The state diagram is split into two parts: the upper half is shown on the left, and the lower half on the right, due to space limitations.
As intended, no states are repeated.

\section{Related Work}

Execution traces are heavily used in time-travel debugging.
Execution of an implemented program in general requires a large number of execution steps, producing a massive number of states in various layers in the program, such as application code, runtime libraries, and so on. 
Willembrinck et al. proposed Time-Traveling Queries (TTQs) to address the scalability issue\cite{TimeTravelingQueries}.
Their approach is to use queries to selectively collect only meaningful execution states that match the given query, such as dispatches of a certain method, read accesses to a specified variable, and write accesses to any variables in a specified method.
Our implementation does not employ selective collection of traces.
Since  VDM-SL is a formal modeling language, all stateful elements are explicitly declared as state variables so that the specified system is free from hidden states such as those in standard libraries.

Regarding visualization, Lausdahl et al.~\cite{FM09} introduced transformations between VDM++ source code and UML diagrams.
Their approach is based on transformations over abstract syntax trees (ASTs). Specifically, they proposed transformations from VDM++ ASTs to UML class diagrams, as well as from VDM++ trace definitions to UML sequence diagrams.
Note that the trace definition in VDM++ is a language feature of the VDM family for specifying combinatorial tests, and is therefore distinct from the execution traces discussed in this paper.

Lund et al.~\cite{Lund} also presented a transformation from VDM++ source code to UML class diagrams.
Their tool, integrated into the Overture environment in Visual Studio Code, generates class diagrams in PlantUML format from VDM++ source code, representing the static structure of the model from an object-oriented perspective.
In contrast, ViennaTalk focuses on the dynamic behavior of the system.
It generates sequence diagrams and state diagrams in Mermaid notation from the execution traces of VDM-SL expressions or statements, thereby capturing the concrete dynamics of a specific execution.
 
\section{Conclusion and Future Work}

Recording execution traces reifies transient states and operation activations into persistent objects.
We have successfully implemented a feature to record execution traces and visualize nested operation calls as sequence diagrams and state transitions as state diagrams.

A promising next step is the design of a query language to express and extract specific patterns from a series of events. 
A significant advantage of utilizing execution traces, as opposed to standard VDM-SL assertions, is the ability to express properties across an arbitrary number of execution steps. 
While invariants and preconditions typically express properties at a single execution step, and postconditions cover two steps (the pre-state and post-state), a trace-based query can capture complex temporal behaviors. 
We are currently designing a query language for execution records, which may extend our work into the fields of runtime verification and regression testing across multiple versions of specifications and implementations. 
We consider process algebras, such as CSP~\cite{CSP}, as a promising formalism to express and analyze the dynamic behavior of systems within recorded event series.

\section*{Acknowledgements}
A part of this research was supported by JSPS KAKENHI Grant Number JP 23K01632, 24K09052 and 23K11058, and Nanzan University Pache Research Subsidy I-A-2 for the 2025 academic year.
The authors would also like to thank the anonymous reviewers for their valuable comments and suggestions.
\bibliographystyle{splncs03}

\bibliography{references}

@TECHREPORT{VDM10LRM,
  KEY           = "VDM10LRM",
  AUTHOR        = "Peter Gorm Larsen and Kenneth Lausdahl and Nick Battle and John Fitzgerald and Sune Wolff and Shin Sahara and Marcel Verhoef and Peter W. V. Tran-J{\o}rgensen and Tomohiro Oda",
  TITLE         = "{{VDM-10} Language Manual}",
  YEAR          = "2013",
  NUMBER        = "TR-001"}

@INPROCEEDINGS{VDMUnit,
  AUTHOR        = "Peter W. V. Tran-J\o{}rgensen and Ren\'e Nilsson and Kenneth Lausdahl",
  EDITOR        = "Ken Pierce and Marcel Verhoef",
  TITLE         = "{Enhancing Testing of {VDM-SL} Models}",
  BOOKTITLE     = "The 16th Overture Workshop",
  ORGANIZATION  = "Newcastle University, School of Computing",
  ADDRESS       = "Oxford",
  YEAR          = "2018",
  PAGES         = "7--22"}

@INPROCEEDINGS{CombinatorialTesting,
  AUTHOR	= "Peter Gorm Larsen and Kenneth Lausdahl and Nick Battle",
  TITLE		= "{Combinatorial Testing for {VDM}}",
 booktitle 		= {Proceedings of the 2010 8th IEEE International Conference on Software Engineering and Formal Methods},
 series 		= {SEFM '10},
 year 		= {2010},
 NOTE 		= "{ISBN 978-0-7695-4153-2}",
 pages 		= {278--285},
 numpages 	= {8},
 url 			= {http://dx.doi.org/10.1109/SEFM.2010.32},
 publisher 	= {IEEE Computer Society},
 address 		= {Washington, DC, USA},
 doi		= {10.1109/SEFM.2010.32},
}

@INPROCEEDINGS{WiderRangeOfStakeholders,
  AUTHOR        = "Tomohiro Oda and Yasuhiro Yamomoto and Kumiyo Nakakoji and
                   Kenjiro Araki and Peter Gorm Larsen",
  EDITOR        = "Fuyuki Ishikawa and Peter Gorm Larsen",
  TITLE         = "{{VDM} Animation for a Wider Range of Stakeholders}",
  BOOKTITLE     = "Proceedings of the 13th Overture Workshop",
  PUBLISHER     = "Center for Global Research in Advanced Software Science
                   and Engineering",
  ADDRESS       = "National Institute of Informatics, 2-1-2 Hitotsubashi,
                   Chiyoda-Ku, Tokyo, Japan",
  YEAR          = "2015",
  PAGES         = "18-32"}

@INPROCEEDINGS{CodeGenerator,
  AUTHOR        = "Tomohiro Oda and Keijiro Araki and Peter Gorm Larsen",
  TITLE         = "{Automated {VDM-SL} to Smalltalk Code Generators for Exploratory Modeling}",
  BOOKTITLE     = "The 14th Overture Workshop: Towards Analytical Tool Chains",
  ORGANIZATION  = "Aarhus University, Department of Engineering",
  EDITOR        = "Peter Gorm Larsen and Nico Plat and Nick Battle",
  PUBLISHER     = "Aarhus University, Department of Engineering",
  ADDRESS       = "Cyprus",
  YEAR          = "2016",
  PAGES         = "48--62"}

@ARTICLE{ViennaTalk,
  AUTHOR        = "Tomohiro Oda and Keijiro Araki and Peter Gorm Larsen",
  TITLE         = "A Formal Modeling Tool for Exploratory Modeling in Software
                   Development",
  JOURNAL       = "IEICE Transactions on Information and Systems",
  YEAR          = "2017",
  VOLUME        = "100",
  NUMBER        = "6",
  PAGES         = {1210--1217}}

@inproceedings{Lund,
  title={Towards {UML} and {VDM} support in the {VS Code} environment},
  author={Lund, Jonas and Jensen, Lucas Bjarke and Macedo, Hugo Daniel and Larsen, Peter Gorm},
  booktitle={Proceedings of the 20th Overture Workshop},
  pages={50--65},
  year={2022}
}

@inproceedings{FM09,
  title={Connecting {UML} and {VDM++} with open tool support},
  author={Lausdahl, Kenneth and Lintrup, Hans Kristian Agerlund and Larsen, Peter Gorm},
  booktitle={International Symposium on Formal Methods},
  pages={563--578},
  year={2009},
  organization={Springer}
}

@inproceedings{TimeTravelingQueries,
  title={Time-traveling debugging queries: Faster program exploration},
  author={Willembrinck, Maximilian and Costiou, Steven and Etien, Anne and Ducasse, St{\'e}phane},
  booktitle={2021 IEEE 21st International Conference on Software Quality, Reliability and Security (QRS)},
  pages={642--653},
  year={2021},
  organization={IEEE}
}

@inproceedings{RefactoringBrowser,
  title={Refactoring for Exploratory Specification in {VDM-SL}},
  author={Oda, Tomohiro and Araki, Keijiro and Sahara, Shin and Chang, Han-Myung and Gorm, Peter},
  booktitle={Proceedings of the 19th Overture Workshop},
  pages={21--35},
  year={2021}
}

@book{Pharo,
  author = {Andrew P. Black and St\'ephane Ducasse and Oscar Nierstrasz and Damien Pollet and Damien Cassou and Marcus Denker},
  title = {Pharo by Example},
  pages = {333},
  publisher = {Square Bracket Associates},
  year = {2009},
  isbn = {978-3-9523341-4-0}
}

@book{CSP,
  title={Understanding concurrent systems},
  author={Roscoe, Andrew W},
  year={2010},
  publisher={Springer Science \& Business Media}
}

@book{Mermaid,
  title={The official guide to Mermaid. js: create complex diagrams and beautiful flowcharts easily using text and code},
  author={Sveidqvist, Knut and Jain, Ashish},
  year={2021},
  publisher={Packt Publishing Ltd}
}

\end{document}